\documentclass[12pt,epsfig]{article}

\usepackage{amssymb,epsfig}
\usepackage[hypertex]{hyperref}

\newcommand{\be}{\begin{equation}}

\newcommand{\ee}{\end{equation}}

\newcommand{\bea}{\begin{eqnarray}}

\newcommand{\eea}{\end{eqnarray}}

\begin{document}
\begin{titlepage}

\begin{flushright}
\begin{tabular}{l}
 CPHT-RR028.0413
\end{tabular}
\end{flushright}
\vspace{1.5cm}

\begin{center}
{\LARGE \bf
QCD description of charmonium plus light meson production in
$\bar{p} -N$
annihilation}
\vspace{1cm}

\renewcommand{\thefootnote}{\alph{footnote}}

{\sc B.~Pire}${}^{1}$,
{\sc K.~Semenov-Tian-Shansky}${}^2$%
\footnote{e-mail:\texttt{kirill.semenov@polytechnique.edu} },
{\sc L.~Szymanowski}${}^{3}$.
\\[0.5cm]
\vspace*{0.1cm} ${}^1${\it
CPhT, {\'E}cole Polytechnique, CNRS, F-91128 Palaiseau, France
                       } \\[0.2cm]
 \vspace*{0.1cm} ${}^2${\it
 IFPA, d\'{e}partement AGO,  Universit\'{e} de  Li\`{e}ge, 4000 Li\`{e}ge,  Belgium
                       } \\[0.2cm]
\vspace*{0.1cm} ${}^3$ {\it
 National Centre for Nuclear Research (NCBJ), Warsaw, Poland
                       } \\[1.0cm]
{\it \large
\today
 }
\vskip2cm
{\bf Abstract:\\[10pt]} \parbox[t]{\textwidth}{
  The associated production of a $J/\psi$ and a $\pi$ meson in antiproton-nucleon annihilation  is studied in the
  framework of QCD collinear factorization. In this approach, a hard subprocess responsible for the production of the
  heavy quark-antiquark pair factorizes   from soft hadronic matrix elements, such as the antiproton (nucleon)
  distribution amplitude and the nucleon-to-pion (antiproton-to-pion) transition distribution amplitude. This reaction
  mechanism should dominate the forward and backward kinematical regions, where the cross sections are expected to
  be measurable in the set-up of the  \={P}ANDA experiment at the GSI-Fair facility.  }
\vskip1cm
\end{center}

\vspace*{1cm}
\end{titlepage}

\section{Introduction}
\setcounter{footnote}{0}

The factorization of exclusive amplitudes in a short distance dominated part which may be calculated in a
perturbative way in the one hand, and universal confinement related hadronic matrix elements in the other hand,
is a welcome feature of quantum chromodynamics (QCD) for specific processes in specific kinematics. The textbook
example of such a  factorization is nearly forward deeply virtual Compton scattering where generalized parton
distributions (GPDs) are the relevant hadronic matrix elements. The extension of this description to other processes such
as backward virtual Compton scattering and backward meson electroproduction, has been advocated
\cite{Frankfurt:1999fp,Pire:2004ie,Pire:2005ax}
 - although not proven. In the latter process, new hadronic matrix elements of three quark operators on the light cone,
the nucleon-to-meson transition distribution amplitudes (TDAs), appear which shed a new light on the nucleon structure,
and enable to quantify nucleon's mesonic cloud. The validity of such a factorization requires at least the existence of
a large scale
$Q$,
which has been taken  as the spacelike, respectively   timelike,  virtuality of the photon quantifying the electromagnetic
probe in the case of electroproduction, respectively  lepton pair emission in
antinucleon-nucleon annihilation. This large
scale ensures the validity of the perturbative expansion of the hard subprocess with the QCD coupling constant
$\alpha_s(\mu)$
taken at a scale
$\mu =O(Q)$.
The factorization scale is also to be taken of order
$Q$.

In many instances (jet production, heavy meson production or decays...)  it has been shown that the occurrence of a
heavy elementary particle, for instance a quark with mass
$m_Q$,
is sufficient to ensure the reliability of a perturbative expansion with
$\mu= m_Q$.
The pioneering studies of the charmonium decay width and of specific charmonium decay channels
\cite{Lepage:1980,Chernyak:1983ej,Chernyak_Nucleon_wave} have proven the value
of this approach. We will follow this line of reasoning in the present Letter.

The spectroscopy of charmonium states is at the heart of the physics program of the \={P}ANDA experiment at the
GSI-Fair facility
\cite{Lutz:2009ff, Wiedner:2011mf}.
Some of these unknown (or badly known) states will decay in a lower mass charmonium such as the
$J/\psi$
and a few ordinary mesons. When scrutinizing the final states produced, it will be of the utmost importance to
separate these associated charmonium-light meson states from a background where the light meson(s) just
evaporate from the beam or the target nucleon. These latter cases are the target of our study. But we stress that
their production process is much more interesting than a mere background to subtract: it is indeed a new way
to access the inner structure of the nucleon through the study of baryon to meson transition distribution amplitudes.
For definiteness, we will present the study in the case of a single
$\pi$
meson produced in conjunction with the
$J/\psi$.
A slight extension of the same formalism applies  in the cases where another meson,
{\it e.g.} $\rho, \eta, \omega, f_0, \varphi $
or a pair of
 mesons, {\it  e.g.}
 $\pi \pi, K K$,
 is produced.

An alternative description  of
$\bar N N$
annihilation in a charmonium accompanied by a  meson proposed in the literature
is the use of effective hadron exchange models; see e.g.
\cite{Gaillard:1982zm} and  \cite{Barnes:2010yb,Lin:2012ru}.
This method provides an essentially non-perturbative description of the process. Our aim is to single
out with the help of factorization techniques the perturbative part of the process and to relate the remaining
non-perturbative part to universal fundamental quantities with interpretation within QCD as the light-cone
matrix elements of correlators of the fundamental fields.

\section{Kinematics}
The study of  charmonium exclusive hadronic decays has  been for a long time one of the fields of application
of perturbative QCD methods.
It has been argued
\cite{Brodsky:1981kj,Chernyak:1983ej}
that the dominating mechanism is the
$c \bar{c}$
pair annihilation into the minimal possible number of gluons which then produce quark-antiquark
pairs forming the outgoing hadrons.

\begin{figure}[t]
\centerline{\epsfxsize6.5cm\epsffile{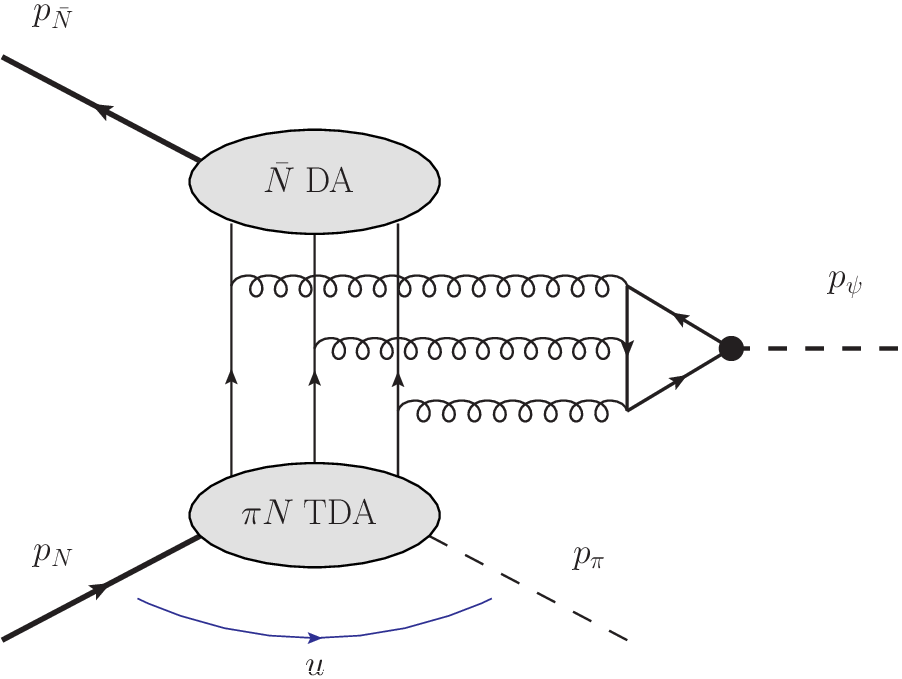} \epsfxsize6.5cm\epsffile{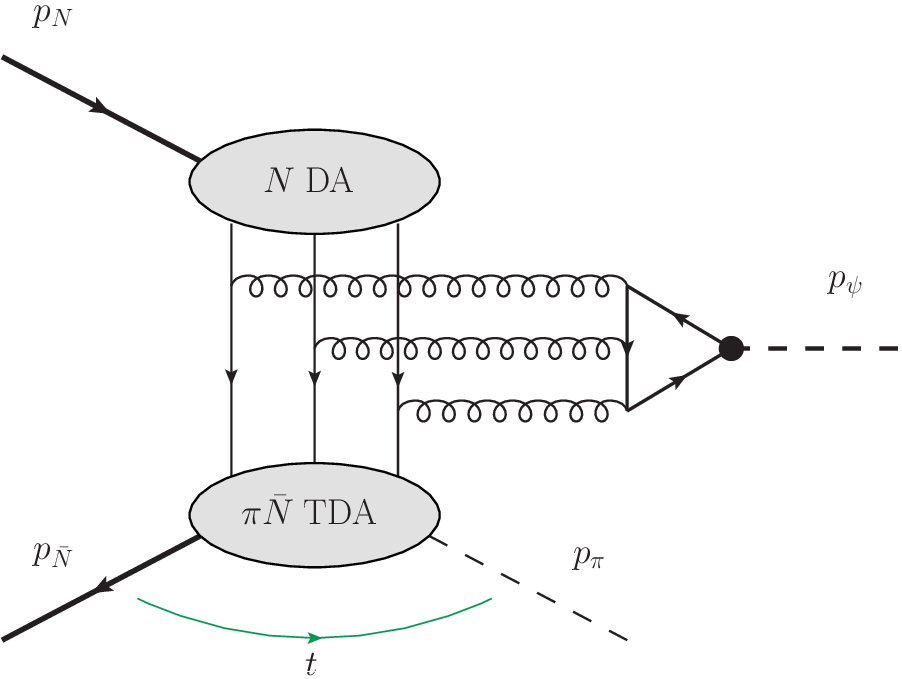} }
\caption[]{\small
Collinear factorization of the annihilation process
$\bar N (p_{\bar N}) N (p_N) \to  J/\psi(p_{\psi}) \pi(p_{\pi})$.
{\bf Left panel:} backward kinematics
($u \sim 0$).
{\bf Right panel:} forward kinematics
($t \sim 0$).
$\bar{N}(N)$
DA stands for the distribution amplitude of antinucleon (nucleon);
$\pi N (\pi \bar{N})$
TDA stands for the transition distribution amplitude from a nucleon (antinucleon) to a pion.}
\label{Fig_factorization}
\end{figure}

In the present Letter we extend the same perturbative QCD framework for the description of a cross
channel reaction in which nucleon-antinucleon annihilate producing heavy quarkonium together with
a light meson
($\pi$, $\eta$, $\omega$, $\rho$)
almost collinear either with incoming nucleon or antinucleon. For simplicity below we address the case of
$\pi$
meson and consider the reaction
\be
\label{reac}
 N (p_N) \;+ \bar N (p_{\bar N}) \; \to  J/\psi(p_{\psi})\;+\; \pi(p_{\pi}).
\ee
Here the
$N \bar{N}$
center-of-mass energy squared
$s=(p_N+p_{\bar{N}})^2 \equiv W^2$
and the charmonium mass squared
$M^2_\psi$
introduce the natural hard scale. In the complete analogy with our analysis
\cite{Lansberg:2007se,Lansberg:2012ha}
we assume that this reaction admits a factorized description  within two distinct
kinematical regimes
(see Fig.~\ref{Fig_factorization}):
\begin{itemize}
\item the  near-forward kinematics
$t \equiv (p_\pi-p_{\bar{N}})^2 \sim 0$;
it corresponds to the pion moving almost in the direction of the initial antinucleon in
$N \bar{N}$
center-of-mass system (CMS);
\item the near-backward kinematics
$u \equiv (p_\pi-p_{N})^2 \sim 0$
corresponding to the pion moving almost in the direction of the initial  nucleon in
$N \bar{N}$
CMS.
\end{itemize}
The suggested reaction mechanism forms the distinct forward and backward peaks of the
differential cross section of the reaction
(\ref{reac})
$d \sigma/ dt$
as the function of
$\cos \theta^*_\pi$ ($\theta^*_\pi$
is the pion scattering angle in the
$N \bar{N}$ CMS).
The process
(\ref{reac})
allows us to test the universality of
$\pi N$
TDAs that appear also in the description of
$\gamma^* N \to \pi N$
and
$N \bar{N} \to \ell^+ \ell^- \pi$
reactions
\cite{Lansberg:2011aa,Lansberg:2012ha}.

Due to the  $C$-invariance of strong interaction there exists a perfect symmetry between the forward and backward
kinematics regimes of the reaction
(\ref{reac}).
These two regimes can be considered in exactly  the same way (see the discussion in Appendix~C of
Ref.~\cite{Lansberg:2012ha}).
Precisely, the amplitude of the reaction
(\ref{reac})
within the $t$-channel factorization regime can be obtained from that  within the $u$-channel
factorization regime with the obvious change of the kinematical variables:
\bea
&&
p_N  \rightarrow p_{\bar{N}};  \ \ \   p_{\bar{N}}  \rightarrow p_{{N}}; \nonumber \\ &&
\Delta^u \equiv (p_\pi-p_N)  \rightarrow \Delta^t \equiv (p_\pi-p_{\bar N});  \nonumber \\ &&
u  \rightarrow t.
\label{tu_change}
\eea

However,  the backward  and the forward regimes are treated somewhat unequally within the  \={P}ANDA experimental set-up
operating antiproton beam. Indeed, once switching to the laboratory system (which corresponds to the nucleon at rest)
one may check that  the forward peak of the cross section as the function of
$\cos \theta_\pi^{\rm LAB}$
is narrowed, while  the backward peak is broadened  by the effect of the Lorentz boost from the
$N \bar{N}$
CMS to the laboratory frame.

In this Letter we have chosen to present explicitly  the details of calculation of the reaction amplitude  within
the backward kinematics regime. As usual, the $z$ axis is chosen along the colliding nucleon-antinucleon with
the positive direction defined by that of the antinucleon beam. We introduce the  light-cone vectors
$p, n$
satisfying
$ 2p \cdot n =1$.
The Sudakov decomposition of the relevant momenta  is presented in the Appendix A of
\cite{Lansberg:2012ha}.

For the calculation of the hard part of the amplitude  we apply the collinear approximation.  We neglect  both the
nucleon and pion masses and assume
$\Delta_T=(p_\pi-p_N)_T=0$,
where the transverse direction is defined with respect to the $z$ direction.

The Sudakov decomposition employed  for the calculation of the hard part then reads:
\bea
&&
p_N \simeq (1+\xi) p ; \ \ \ \
p_{\bar{N}} \simeq \frac{s}{(1+\xi)} n;   \nonumber \\ &&
p_\psi \simeq 2 \xi p + \frac{s}{(1+\xi)} n ; \ \ \ \ \Delta  \equiv \Delta^u = p_\pi-p_N \simeq -2 \xi p .
\eea
Here
$\xi$ is the $u$-channel skewness variable
\be
\xi \equiv - \frac{(p_\pi-p_N) \cdot n}{(p_\pi+p_N) \cdot n} \simeq \frac{M^2_\psi}{2 W^2 -M^2_\psi}.
\ee
Following
\cite{Brodsky:1981kj},
in our calculation we set
\be
M_\psi  \simeq  2m_c   \simeq \bar{M} ,
\label{mass approx}
\ee
taking the average value
$\bar{M}= 3 \; {\rm GeV }$.

Let us emphasize that we keep the exact kinematics (see Appendix.~A of Ref.
\cite{Lansberg:2012ha})
for the nucleon spinors and for $J/\psi$ polarization vector. The physical kinematical domain for the reaction
(\ref{reac})
in the backward regime is determined by the requirement
${\Delta_T^2} \le 0$,
where
\bea
{\Delta_T^2}= \frac{1-\xi}{1+\xi} \left(\Delta^2- 2\xi \left[\frac{m_N^2}{1+\xi} -\frac{m_\pi^2}{1-\xi} \right] \right),
\eea
where $m_N$ ($m_\pi$) stands for the nucleon (pion) mass.
The limiting value
${\Delta_T^2}=0$
corresponds to
\be
\Delta^2=\Delta^2_{\max} \equiv \frac{2 \xi  \left( {m_N^2} (\xi -1)+m_\pi ^2 (\xi +1)\right)}{\xi ^2-1}.
\label{Dmax}
\ee
In the calculations presented below we neglect the pion mass
$m_\pi$.

\section{Hard amplitude calculation  for $N + \bar{N} \to  J/\psi + \pi  $}

\begin{figure}[h]
 \begin{center}
 \epsfig{figure= 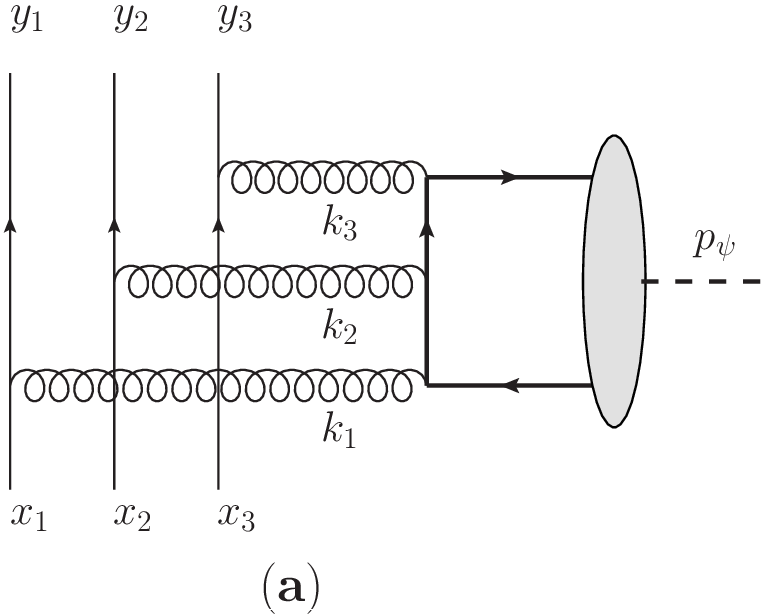  , height=4.0cm} \ \
 \epsfig{figure=  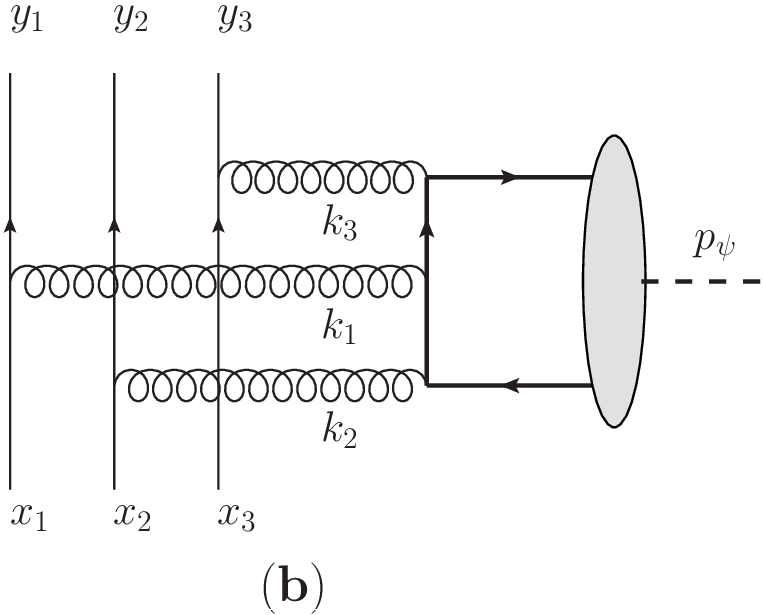 , height=4.0cm} \ \
  \epsfig{figure=  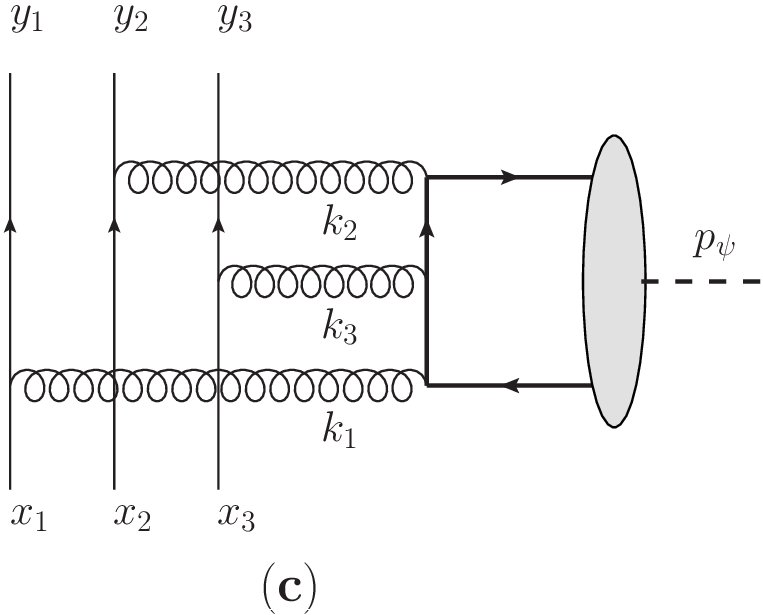 , height=4.0cm}
   \end{center}
     \caption{ Feynman diagrams describing $J/\psi ~\pi$ production subprocess at the Born order. }
\label{Fig_diagrams}
\end{figure}

The calculation of
$N + \bar{N} \to J/\psi+  \pi $
scattering amplitude follows the same main steps as the classical calculation of the
$J/\psi \to p + \bar{p}$ amplitude
\cite{Brodsky:1981kj,Chernyak:1983ej,Chernyak:1987nv}.
Assuming the factorization of small and large distance dynamics the hard part of the amplitude is computed within
perturbative QCD. Large distance dynamics is encoded within the matrix elements of QCD light-cone operators between
the appropriate hadronic states. The leading order amplitude of
(\ref{reac})
is then given by the sum of the three diagrams presented on Figure~\ref{Fig_diagrams}.

Below we present our conventions for the relevant light-cone matrix elements encoding the soft dynamics.
For definiteness we consider the case of the leading twist
$uud$ $p \pi^0$
TDA. The case of
$n \pi^-$
TDA is completely analogous. Throughout this Letter we make use of the parametrization of
\cite{Lansberg:2007ec}%
\footnote{The relation between this parametrization and the one
employed in Refs. \cite{Pire:2011xv,Lansberg:2011aa,Lansberg:2012ha} is given in the Appendix A of \cite{Lansberg:2011aa}. } since it
simplifies considerably for
$\Delta_T=0$
case.
The parametrization involves
$8$
invariant functions each being the function of three longitudinal momentum fractions
$x_i$,
skewness variable
$\xi$,
momentum transfer squared
$\Delta^2$
as well as of the factorization scale
$\mu$:
\bea
&&
4 (p \cdot n)^3 \int \left[ \prod_{j=1}^3 \frac{d \lambda_j}{2 \pi}\right]
e^{i \sum_{k=1}^3 x_k \lambda_k (p \cdot n)}
 \langle     \pi^0(p_\pi)|\,  \varepsilon_{c_1 c_2 c_3} u^{c_1}_{\rho}(\lambda_1 n)
u^{c_2}_{\tau}(\lambda_2 n)d^{c_3}_{\chi}(\lambda_3 n)
\,|N^p(p_N,s_N) \rangle
\nonumber \\ &&
= \delta(x_1+x_2+x_3-2 \xi) i \frac{f_N}{f_\pi}\Big[  V^{(p\pi^0)}_{1}(x_{1,2,3}, \xi,\Delta^2)  (  \hat{p} C)_{\rho \tau}(U^+)_{\chi}
\nonumber \\ &&
+A^{(p\pi^0)}_{1}(x_{1,2,3}, \xi,\Delta^2)  (  \hat{p} \gamma^5 C)_{\rho \tau}(\gamma^5 U^+ )_{\chi}
 +T^{(p\pi^0)}_{1}(x_{1,2,3}, \xi,\Delta^2)  (\sigma_{p\mu} C)_{\rho \tau }(\gamma^\mu U^+ )_{\chi}
 \nonumber \\ &&
 + m_N^{-1} V^{(p\pi^0)}_{2} (x_{1,2,3}, \xi,\Delta^2)
 ( \hat{ p}  C)_{\rho \tau}( \hat{\Delta}_T U^+)_{\chi}
+ m_N^{-1}
 A^{(p\pi^0)}_{2}(x_{1,2,3}, \xi,\Delta^2)  ( \hat{ p}  \gamma^5 C)_{\rho\tau}(\gamma^5  \hat{\Delta}_T  U^+)_{\chi}
  \nonumber \\ &&
+ m_N^{-1} T^{(p\pi^0)}_{2} (x_{1,2,3}, \xi,\Delta^2) ( \sigma_{p\Delta_T} C)_{\rho \tau} (U^+)_{\chi}
+  m_N ^{-1}T^{(p\pi^0)}_{3} (x_{1,2,3}, \xi,\Delta^2) ( \sigma_{p\mu} C)_{\rho \tau} (\sigma^{\mu\Delta_T}
 U^+)_{\chi}
 \nonumber \\ &&
+ m_N^{-2} T^{(p\pi^0)}_{4} (x_{1,2,3}, \xi,\Delta^2)  (\sigma_{p \Delta_T} C)_{\rho \tau}
(\hat{ \Delta}_T U^+)_{\chi} \Big].
 \label{Old_param_TDAs}
\eea
Here
$f_\pi=93$ MeV
is the pion weak decay constant and
$f_N$
determines the value of the nucleon wave function at the origin.
Throughout this paper we adopt Dirac's ``hat'' notation $\hat{v} \equiv v_\mu \gamma^\mu$;
$\sigma^{\mu\nu}= \frac{1}{2} [\gamma^\mu, \gamma^\nu]$; $\sigma^{v \mu} \equiv v_\lambda \sigma^{\lambda \mu}$;
$C$
is the charge conjugation matrix and
$U^+= \hat{p} \hat{n} \, U(p_N,s_N)$
is the large component of the nucleon spinor.
For
$\Delta_T=0$
just three invariant amplitudes
$V^{(p\pi^0)}_{1}$, $A^{(p\pi^0)}_{1}$ and $T^{(p\pi^0)}_{1}$
survive in the parametrization (\ref{Old_param_TDAs}).

For the leading twist antinucleon DAs we employ the standard parametrization
\cite{Chernyak:1983ej}
(see also Appendix~B of Ref.~\cite{Lansberg:2012ha}).
The non-relativistic light-cone wave function of
$J/\psi$
heavy quarkonium is given by
\cite{Chernyak:1983ej}
\bea
\Phi_{\rho \tau}(z, p_\psi)= \langle 0| \bar{c}_\tau(z) c_\rho(-z) | J/\psi \rangle
=
\frac{1}{4} f_\psi
\left[ 2 m_c \hat{\cal E} +\sigma_{ p_{\psi} \nu}   {\cal E}^\nu
\right]_{\rho \tau},
\label{WFNR}
\eea
where $m_c$ is the $c$-quark mass and
${\cal E}$
stands for the charmonium polarization vector. With the use of the non-relativistic wave function
(\ref{WFNR})
we tacitly assume that each charm quark carries half of the momentum of the
$J/\psi$.
The normalization constant
$f_\psi$
is extracted from the charmonium leptonic decay width
$\Gamma(J/\psi \to e^+  e^- )$:
\be
\Gamma(J/\psi \to e^+  e^- )= (4 \pi \alpha_{\rm e.m.})^2 \frac{e_c^2}{12 \pi } f_\psi^2 \frac{1}{M_\psi}, \ \ \ e_c=\frac{2}{3}.
\ee
Using the values quoted in
\cite{PDG2012}
we get
\be
|f_\psi|= 413 \pm 8 \; {\rm MeV}.
\label{fpsi_value}
\ee

The leading order amplitude of
(\ref{reac})
reads
\bea
&&
{\cal M}^{s_N s_{\bar{N}}}_\lambda= {\cal C} \frac{1}{{\bar M}^5 } \Big[
\bar{V}(p_{\bar{N}},s_{\bar{N}} )\hat{\cal E}^*(\lambda) \gamma_5 U(p_N, s_N) {\cal I}(\xi,\Delta^2) \nonumber \\ &&
-\frac{1}{m_N}  \bar{V}(p_{\bar{N}},s_{\bar{N}} )\hat{\cal E}^*(\lambda) \hat{\Delta}_T \gamma_5 U(p_N, s_N) {\cal I}'(\xi,\Delta^2)
\Big],
\label{Amplitude_master}
\eea
where
$\bar{V}$
and
$U$
stand  for the nucleon Dirac spinors. The calculation of $3$ diagrams presented on
Fig.~\ref{Fig_diagrams}
yields the following result for
${\cal I}(\xi,\Delta^2)$:
\bea
&&
\mathcal{I}(\xi,\Delta^2)
 \equiv { {\int^{1+\xi}_{-1+\xi} }\! \! \!
d_3 x   
\, \delta(\sum_{j=1}^3 x_j-2\xi)
}
\;
{{\int^{1}_{0} }\! \! \! d_3y   
\,
\delta(\sum_{k=1}^3 y_k-1) } \nonumber
\\ &&
\left\{ \frac{\xi ^3 (x_1 y_3+x_3 y_1) (V_1^{(\pi N)}(x_{1,2,3},\xi,\Delta^2)-A_1^{(\pi N)}(x_{1,2,3},\xi,\Delta^2))
(V^{p}(y_{1,2,3})-A^{p}(y_{1,2,3})) }{  y_1 y_2 y_3 (x_1+i0) (x_2+i0) (x_3+i0)   (x_1 (2
   y_1-1)-2 \xi  y_1+i0) (x_3 (2 y_3-1)-2 \xi  y_3+i0)} \right.
\nonumber \\ &&
+\left. \frac{ \xi ^3 (x_1 y_2+x_2 y_1) (2 T_1^{(\pi N)}(x_{1,2,3},\xi,\Delta^2)+
\frac{\Delta_T^2}{m_N^2}T_4^{(\pi N)}(x_{1,2,3},\xi,\Delta^2)) T^{p}(y_{1,2,3})}{  y_1 y_2 y_3 (x_1+i0) (x_2+i0) (x_3+i0)
(x_1 (2y_1-1)-2 \xi  y_1+i0) (x_2 (2 y_2-1)-2 \xi  y_2+i0)}  \right\},
   \nonumber \\ &&
   \label{Amplitude_result}
\eea
where
\bea
 {\int^{1+\xi}_{-1+\xi} }\! \! \!
d_3 x   \equiv  {\int^{1+\xi}_{-1+\xi} }\! \! \!
d x_1  {\int^{1+\xi}_{-1+\xi} }\! \! \!
d x_2  {\int^{1+\xi}_{-1+\xi} }\! \! \!
d x_3 ; \ \  \ \ \ \ 
{\int^{1}_{0} } \! \! \! d_3y \equiv {\int^{1}_{0} } dy_1 {\int^{1}_{0} } dy_2 {\int^{1}_{0} } dy_3.
\eea
For
${\cal I}'(\xi,\Delta^2)$
we get
\bea
&&
\mathcal{I}'(\xi,\Delta^2)  
  \equiv { {\int^{1+\xi}_{-1+\xi} }\! \! \!
d_3 x   
\, \delta(\sum_{j=1}^3 x_j-2\xi)
}
%
\;
{{\int^{1}_{0} }\! \! \! d_3y   
\,
\delta(\sum_{k=1}^3 y_k-1) } \nonumber \\ &&
\left\{ \frac{\xi ^3 (x_1 y_3+x_3 y_1) (V_2^{(\pi N)}(x_{1,2,3},\xi,\Delta^2)-A_2^{(\pi N)}(x_{1,2,3},\xi,\Delta^2))
(V^{p}(y_{1,2,3})-A^{p}(y_{1,2,3})) }{  y_1 y_2 y_3 (x_1+i0) (x_2+i0) (x_3+i0)   (x_1 (2
   y_1-1)-2 \xi  y_1+i0) (x_3 (2 y_3-1)-2 \xi  y_3+i0)} \right.
\nonumber \\ &&
+\left. \frac{ \xi ^3 (x_1 y_2+x_2 y_1) (T_2^{(\pi N)}(x_{1,2,3},\xi,\Delta^2)+
T_3^{(\pi N)}(x_{1,2,3},\xi,\Delta^2)) T^{p}(y_{1,2,3})}{  y_1 y_2 y_3 (x_1+i0) (x_2+i0) (x_3+i0)
(x_1 (2   y_1-1)-2 \xi  y_1+i0) (x_2 (2 y_2-1)-2 \xi  y_2+i0)}  \right\}.
   \nonumber \\ &&
   \label{Amplitude_result_Ip}
\eea

The overall factor
$\cal C$
in
(\ref{Amplitude_master})
is expressed as:
\bea
{\cal C}= (4 \pi \alpha_s)^3 \frac{f_N^2 f_\psi}{f_\pi}  \underbrace{ \frac{1}{2}}_{J/\psi \; {\rm w.f.} \atop { \rm  normalization} }
\times \,  \underbrace{16}_{\rm Dirac \; trace} \,  \times\underbrace{ \frac{5}{3} \cdot \frac{1}{3} \cdot \frac{1}{(3!)^2}}_{\rm color \; factor}=
(4 \pi \alpha_s)^3 \frac{f_N^2 f_\psi}{f_\pi}  \, \frac{10}{81},
\eea
where
$\alpha_s$
stands for the strong coupling. One may check that the poles in
$x_i$
are indeed located either on the cross over trajectories
$x_i=0$,
which separate the DGLAP-like and ERBL-like support regions of
$\pi N$ TDAs%
\footnote{For the definition of the ERBL-like and DGLAP-like support regions of TDAs see
\cite{Pire:2010if}.},
or within the DGLAP-like support region (for the $y$-dependent poles in $x_i$) as it certainly should be.

The structure of result
(\ref{Amplitude_master}), (\ref{Amplitude_result}), (\ref{Amplitude_result_Ip})
resembles much the well-known expression for
$J/\psi \to \bar{p} p$
decay amplitude
\cite{Chernyak:1987nv}:
\bea
{\cal M}= (4 \pi \alpha_s)^3 \frac{f_N^2 f_\psi}{ {\bar{M}^5}} \,  \frac{10}{81}  \, \bar{U} \hat{\cal E} V \,  M_0,
\label{Decay_ampl_chernyak}
\eea
where
\bea
&&
M_0=  {\int^{1 }_{0 } }\! \! \!
d_3 x  \delta(\sum_{j=1}^3 x_j-1)
 {\int^{1 }_{0 } }\! \! \!
 d_3 y   \delta(\sum_{k=1}^3 y_k-1)
 \\ &&
\left\{ \frac   {y_1 x_3  (V^{p}(x_{1,2,3})-A^{p}(x_{1,2,3}))  (V^{p}(y_{1,2,3})-A^{p}(y_{1,2,3})) }{  y_1 y_2 y_3 \,  x_1 x_2 x_3
(1-(2x_1-1)(2y_1-1))  (1-(2x_3-1)(2y_3-1))} \right.
\nonumber \\ &&
+\left. \frac{2 y_1 x_2 T^{p}(x_{1,2,3}) T^{p}(y_{1,2,3})}{  y_1 y_2 y_3 \, x_1   x_2   x_3   (1-(2x_1-1)(2y_1-1))  (1-(2x_2-1)(2y_2-1))}  \right\}.
   \nonumber \\ &&
   \label{Def_M0}
\eea
The
$J/\psi \to \bar{p} p$
decay amplitude
(\ref{Decay_ampl_chernyak})
results in the following expression for the decay width
 \cite{Chernyak:1987nv}:
\bea
\Gamma(J/\psi \to p \bar{p} )= (\pi \alpha_s)^6 \frac{1280 f_\psi^2 f_N^4 }{243 \pi {\bar{M}^9}} |M_0|^2. 
\label{Charm_dec_width}
\eea

\section{Estimates of the cross section}

The squared amplitude
(\ref{Amplitude_master})
averaged over spins of initial particles reads
\bea
|\bar{{\cal M}}_{\lambda \lambda'}|^2= \frac{1}{4} \sum_{s_N s_{\bar{N}}}
{\cal M}^{s_N s_{\bar{N}}}_\lambda ({\cal M}^{s_N s_{\bar{N}}}_{\lambda'})^*.
\eea
At the leading twist only the transverse polarization of
$J/\psi$
is relevant. To sum over the transverse polarization we employ the relation:
\bea
\sum_{\lambda_T}  {\cal E}^\nu(\lambda) {\cal E^*}^\mu(\lambda)
=-g^{\mu \nu}+\frac{1}{(p \cdot n)}(p^\mu n^\nu+p^\nu n^\mu),
\eea
and get
\bea
|\overline{\mathcal{M}_{T}}|^2 \equiv \sum_{\lambda_T} |\overline{\mathcal{M}_{T T}}|^2 =
\frac{1}{4} |\mathcal{C}|^2 \frac{2(1+\xi)}{\xi {\bar{M}}^8}  \left( |\mathcal{I}(\xi, \Delta^2)|^2 - \frac{\Delta_T^2}{m_N^2} |\mathcal{I}'(\xi, \Delta^2)|^2 \right).
\eea
The leading twist differential cross section of
$N + \bar{N} \to J/\psi + \pi$
then reads
\cite{CORE}
\bea
\frac{d \sigma}{d \Delta^2}= \frac{1}{16 \pi \Lambda^2(s,m_N^2,m_N^2) } |\overline{\mathcal{M}_{T}}|^2,
\label{CS_def_delta2}
\eea
where
$\Lambda(x,y,z)= \sqrt{x^2+y^2+z^2-2xy-2xz-2yz}$.

In order to get a rough estimate of the cross section we use the simple nucleon exchange model for
$\pi N$ TDAs
suggested in
\cite{Lansberg:2011aa}.
We do not expect that the inclusion of the spectral part for
$\pi N$
TDAs
\cite{Pire:2010if}
would be essential to draw a conclusion on  the feasibility of the relevant experiment and may be postponed
until the precise experimental data will be available.

For the
$\pi N$
TDAs within the parametrization
(\ref{Old_param_TDAs})
the nucleon pole model of Ref.
\cite{Lansberg:2011aa} reads
\bea
&&
\big\{ V_1, \, A_1 , \, T_1  \big\}^{(p \pi^0)} ( 
x_i, \xi,\Delta^2)\Big|_{N(940)}
\nonumber \\ &&
 =\Theta_{\rm ERBL}(x_1,x_2,x_3) \times  (g_{\pi NN}) \frac{m_N f_\pi}{\Delta^2-m_N^2}    \frac{1}{(2 \xi) } \frac{1-\xi}{1+\xi}
 \big\{ V^p,\,A^p, \,T^p  \big\}\left( \frac{x_1}{2 \xi}, \frac{x_2}{2 \xi}, \frac{x_3}{2 \xi} \right);
 \nonumber \\  &&
 \big\{ V_2, \, A_2 , \, T_2, \, T_3  \big\}^{(p \pi^0)} ( 
x_i, \xi,\Delta^2)\Big|_{N(940)}
\nonumber \\ &&
=\Theta_{\rm ERBL}(x_1,x_2,x_3) \times  (g_{\pi NN}) \frac{m_N f_\pi}{\Delta^2-m_N^2}    \frac{1}{(2 \xi) }
 \big\{ V^p,\,A^p, \,T^p, \, T^p  \big\}\left( \frac{x_1}{2 \xi}, \frac{x_2}{2 \xi}, \frac{x_3}{2 \xi} \right);
\nonumber \\  &&
   T_4^{(p \pi^0 )} ( 
x_i, \xi,\Delta^2)\Big|_{N(940)}=0,
\label{Nucleon_exchange_contr_VAT}
\eea
and
\bea
\big\{ V_{1,2}, \, A_{1,2} , \, T_{1,2,3,4}  \big\}^{(n \pi^-)} ( 
x_i, \xi,\Delta^2)\Big|_{N(940)}= \sqrt{2} \big\{V_{1,2}, \, A_{1,2} , \, T_{1,2,3,4}    \big\}^{(p \pi^0 )} ( 
x_i, \xi,\Delta^2)\Big|_{N(940)}.
\nonumber \\ &&
\label{Nucleon_exchange_contr_VAT_pi_minus}
\eea
Here
$V^p$, $A^p$ and $T^p$
stand for the nucleon DAs;
$g_{\pi NN}\approx 13$
is the pion-nucleon phenomenological coupling and
\be
\Theta_{\rm ERBL}(x_1,x_2,x_3)  \equiv  \prod_{k=1}^3 \theta(0 \le x_k \le 2 \xi)
\label{theta_ERBL}
\ee
ensures the pure ERBL-like support of TDAs.
For the simple nucleon pole model
(\ref{Nucleon_exchange_contr_VAT})
\bea
&&
{\cal I}(\xi, \Delta^2)\Big|_{N(940)}=
\frac{  f_\pi \,   g_{\pi NN}  m_N (1-\xi) } {   (\Delta^2-m_N^2) (1+\xi )} M_0;
\nonumber \\ &&
{\cal I}'(\xi, \Delta^2)\Big|_{N(940)}=
\frac{  f_\pi \,   g_{\pi NN}  m_N   } {   (\Delta^2-m_N^2)  } M_0,
\eea
where
$M_0$
is given by eq. (\ref{Def_M0}).

As the phenomenological input for our cross section calculation we may use different solutions for the
leading twist nucleon DA. Similarly to the case of charmonium decay width our result  depends strongly
on the form of input nucleon DA. In order to be able to characterize this dependence we present in
Table~\ref{Table_Jpsiwidth}
the predictions of
$J/\psi$
decay width
(\ref{Charm_dec_width}) into $p \bar{p}$
within the pQCD description with the use of the nucleon DAs in question as the numerical input.
Several points are to be mentioned.
\begin{itemize}
\item The decay width
(\ref{Charm_dec_width})
shows strong dependence on
$\alpha_s$: $\sim \alpha_s^6$.
There is no unique opinion in the literature on the value of the strong coupling for the gluon virtuality in question.
\item Nucleon DAs that are strongly concentrated in the end-point regions such as Chernyak-Ogloblin-Zhitnitsky (COZ)
\cite{Chernyak:1987nv}
or King and Sachrajda (KS)
\cite{King:1986wi}
for
$\alpha_s=0.3$
seem to overestimate the experimental width by the factor
$2 - 4$.
They require  smaller  values of
$\alpha_s \sim 0.25 $
to reproduce the experimental value. These solutions have been strongly criticized in the literature
(see e.g. discussion in Chapter~4 of Ref.~\cite{Brambilla:2004wf}).

\item The
"heterotic"  DA model
\cite{Stefanis_DrNauk}
requires even smaller  values of $\alpha_s \sim 0.2 $ to reproduce the experimental width.

\item On the other hand, the phenomenological solutions for the nucleon DA  which are close to the asymptotic form
(Bolz-Kroll (BK)
\cite{Bolz:1996sw},
Braun-Lenz-Wittmann (BLW NLO)
model of
\cite{Braun:2006hz})
for
$\alpha_s=0.3$
underestimate the experimental value of the decay width by a factor
$5 \div 10$.
Usually,
$\alpha_s \sim 0.4$
is required to reproduce the
experimental value of
$\Gamma(J/\psi  \to p \bar{p})$.
\end{itemize}

\begin{center}
\begin{table}[tbp]
\centering
\begin{tabular}{|l|l|l|l|}
\hline
   &  &   & $\alpha_s$  for  which \\

  DA model  & $|f_N|$ GeV$^{2}$  &$\Gamma(J/\psi \to p \bar{p} )$ (KeV)  &   $\Gamma^{\rm Exp.}(J/\psi \to p \bar{p} )$  \\
     &   & for $\alpha_s=\bar{\alpha_s}=0.3$  &  is reproduced  \\
  \hline
    COZ &  $(5.0 \pm 0.5)  \cdot 10^{-3}$  & 0.42 & 0.26 \\
    KS  &     $(5.0 \pm 0.5)  \cdot 10^{-3}$                              &  0.84 & 0.24 \\
    ``Heterotic'' model  & $(5.0 \pm 0.5)  \cdot 10^{-3}$ & 1.20& 0.22 \\
    BK &  $6.64 \cdot 10^{-3}$ & 0.05 & 0.38\\
    BLW NLO & $(5.0 \pm 0.5)  \cdot 10^{-3}$ & 0.02  & 0.44\\
    Asymptotic &  $(5.0 \pm 0.5)  \cdot 10^{-3}$ & 0.015 & 0.46 \\

\hline
\hline
Experiment & -- & $0.19 - 0.21$ &--\\
\hline
\end{tabular}
\caption{\label{Table_Jpsiwidth}
Nucleon light-cone wave function normalization constant
$f_N$
at
$\mu_0=1$ GeV$^2$
and
$J/\psi \to p \bar{p}$
decay width for various nucleon DA models. Experimental value
is  taken from
\cite{PDG2012}.}
\end{table}
\end{center}

\begin{figure}[h]
 \begin{center}
 \epsfig{figure= 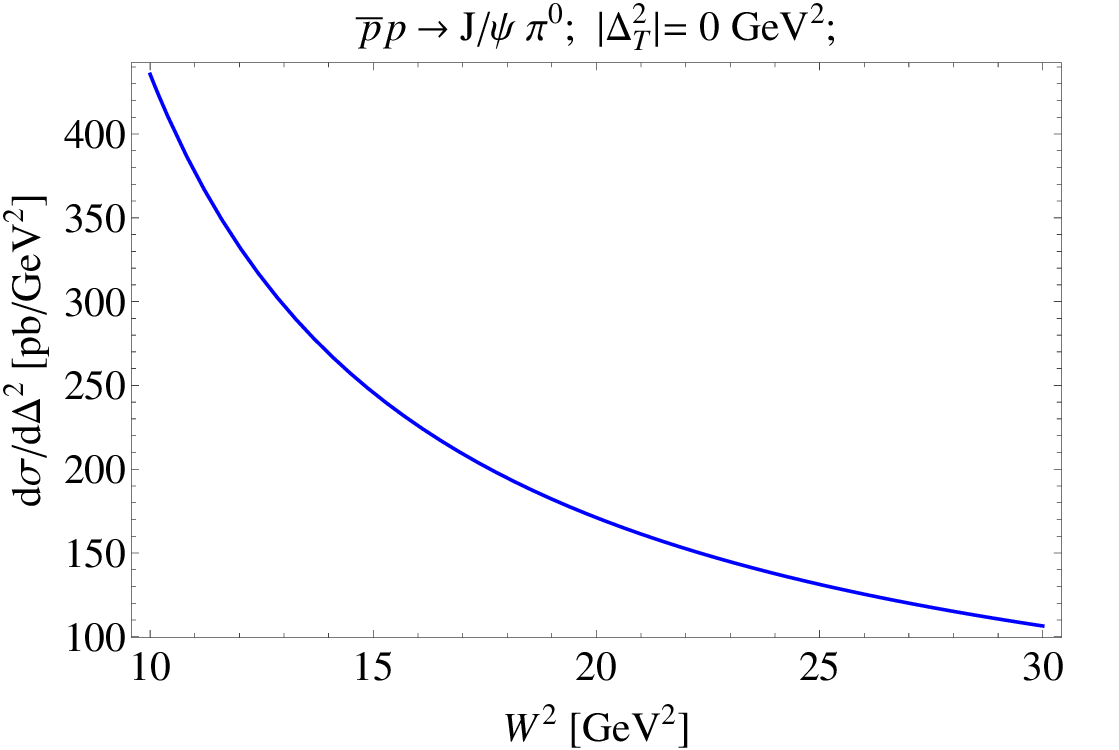 , height=5.0cm}
    \end{center}
     \caption{Differential cross section
$\frac{d \sigma}{d \Delta^2}$ for $p \bar{p} \to J/\psi \,  \pi^0$
as a function of
$W^2$
for
$\Delta_T^2=0$. }
\label{Fig_CS_W2}
\end{figure}

\begin{figure}[h]
 \begin{center}
 \epsfig{figure= 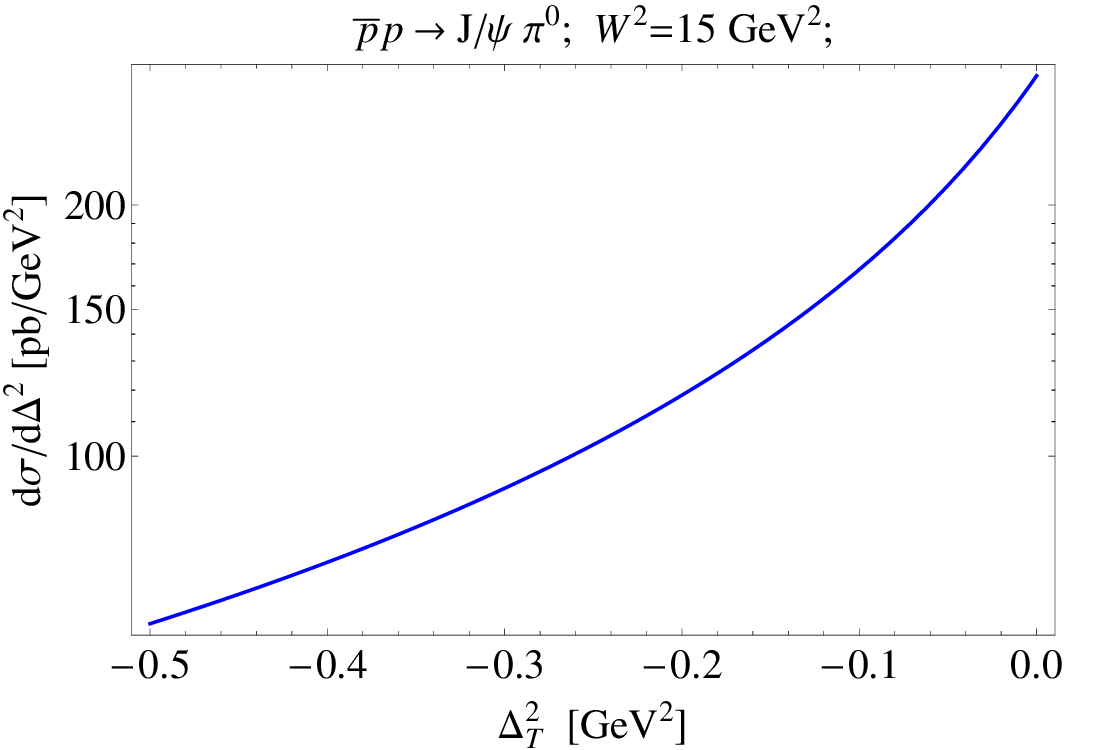, height=5.0cm}
 \epsfig{figure= 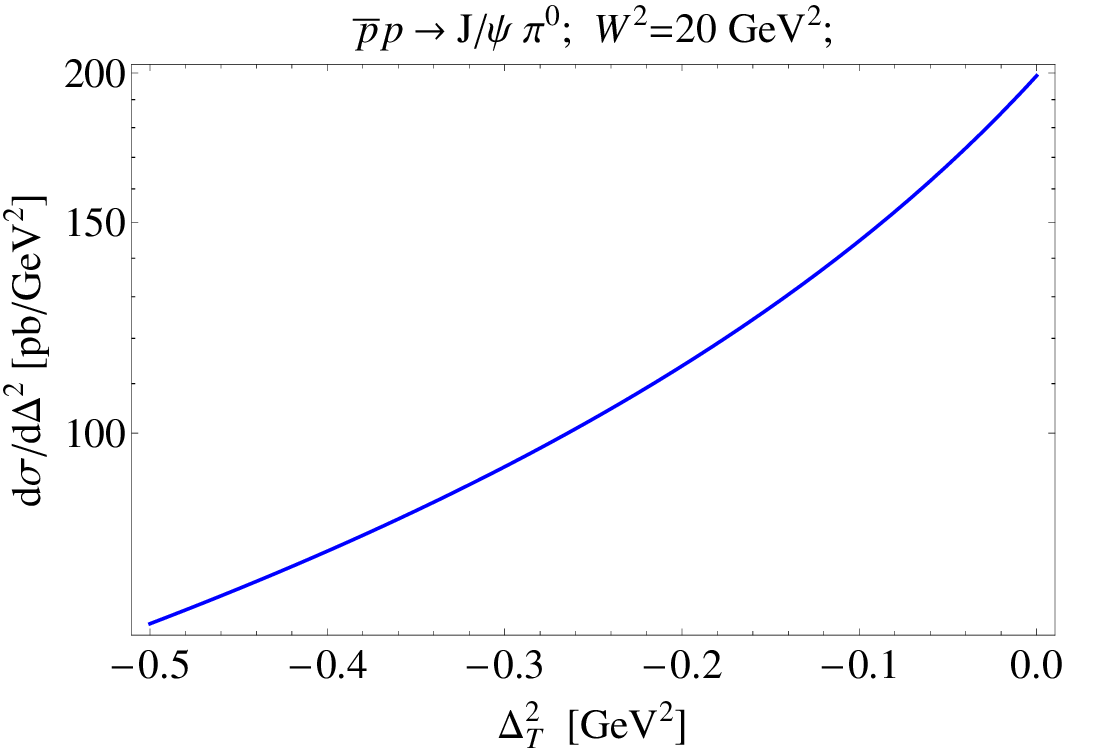, height=5.0cm}
   \end{center}
     \caption{ Differential cross section
$\frac{d \sigma}{d \Delta^2}$ for $p \bar{p} \to J/\psi \,  \pi^0$
as a function of
$\Delta_T^2$
for
$W^2=15$ GeV$^2$
(left panel) and
$W^2=20$ GeV$^2$
(right panel). }
\label{Fig_CS_DeltaT2}
\end{figure}

Based on the above arguments we have chosen to  present our results for the
$N \bar{N} \to J/\psi \,  \pi $
cross section with the value of
$\alpha_s$
fixed by the requirement that the given
phenomenological solution reproduces the experimental
$J/\psi \to N \bar{N}$
decay width.
On Fig.~\ref{Fig_CS_W2}
we show our estimates of the differential cross section
$\frac{d \sigma}{d \Delta^2}$ for $p \bar{p} \to J/\psi \,  \pi^0$
as a function
of
$W^2$
for
$\Delta_T^2=0$.
On Fig.~\ref{Fig_CS_DeltaT2}
we show the differential cross section
$\frac{d \sigma}{d \Delta^2}$
for
$p \bar{p} \to J/\psi \,  \pi^0$
as a function of
$\Delta_T^2$
for several values of
$W^2$.

The pion scattering angle in the
$N \bar{N}$
CMS for the
$u$-channel factorization regime then can be expressed as:
\be
\cos \theta_{\pi}^*=\frac{-(1-\xi )\alpha+\frac{m^2_\pi-{\Delta_T^2} }{1-\xi } \beta }
{\sqrt{(-(1-\xi )\alpha +\frac{m^2_\pi-{\Delta_T}^2}{1-\xi } \beta )^2-{\Delta_T^2}}},
\label{CosThetaPi_uregime}
\ee
where
\be
\alpha=\frac{W+\sqrt{W^2-4 m_N^2}}{4 (1+\xi)}; \ \ \ \beta = \frac{\left(W-\sqrt{W^2-4 m_N^2} \right) (1+\xi)}{4 m_N^2}.
\label{Def_alpha_beta_u}
\ee
One may check that for
${\Delta_T}^2=0$
indeed
$\cos \theta_\pi^*=-1$,
which means backward scattering. Expressing our cross sections as functions of
$\cos \theta_\pi^*$
(\ref{CosThetaPi_uregime})
we reproduce the order of magnitude
($\sim 100 - 300$ pb/GeV$^2$ for $\Delta_T^2=0$)
of the corresponding cross sections presented in
\cite{Lin:2012ru}.
This fact is certainly not surprising since it can be seen as an artefact of a simple nucleon pole model of
$\pi N$
TDAs.

\begin{figure}[h]
 \begin{center}
 \epsfig{figure= 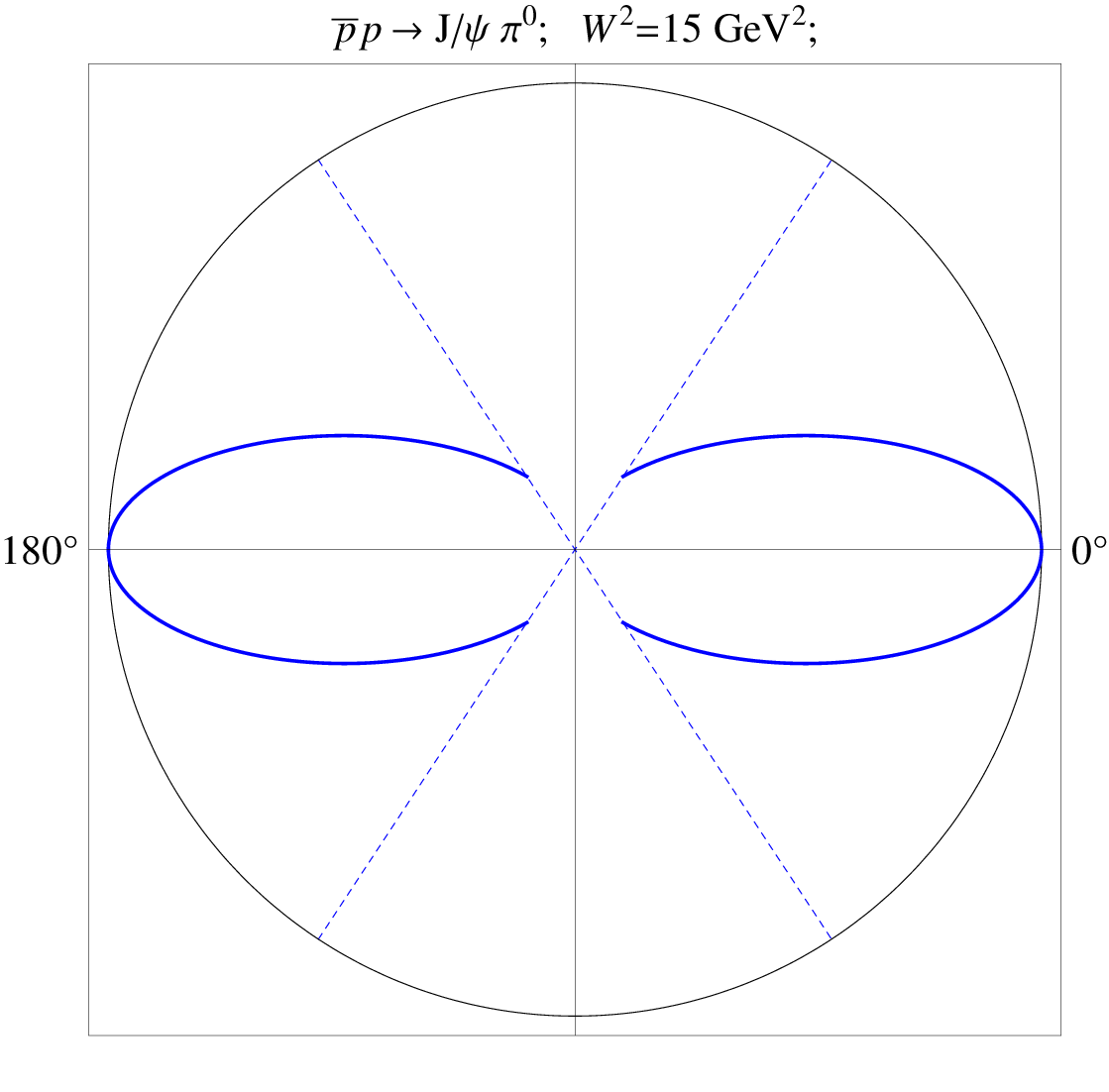 , height=7cm}
 \end{center}
     \caption{ Angular distribution for the
$d \sigma/d \Delta^2$
cross section for near-forward
($\cos \theta_\pi^*>0$)
and near-backward
($\cos \theta_\pi^*<0$) scattering regimes for   $-1 \, {\rm GeV}^2 \le  \Delta^2 \le \Delta^2_{\max}$.
Dashed lines show the effect of the cutoff
$\Delta^2 \ge -1$  GeV$^2$
for the values of the pion CMS scattering angle $\theta_\pi^*$.  }
\label{Fig_CS_angular}
  \end{figure}

We also reproduce the characteristic shape of angular distributions of the  cross sections presented in
\cite{Lin:2012ru}.
On Fig.~\ref{Fig_CS_angular} we show the center of mass angular distribution for the
$d \sigma/d \Delta^2$
cross section for both forward and backward factorization regimes presented on the polar plot with the
polar angle being the pion CMS scattering angle
$\theta_\pi^*$.
We present the ratio
\be
\frac{ \frac{d \sigma} {d \Delta^2} (W^2, \Delta_T^2)} {\frac{d \sigma} {d \Delta^2} (W^2, \Delta_T^2=0)}
\ee
as the function of
$\theta_\pi^*$
showing the result for $W^2=15$ GeV$^2$ and for $   -1 \, {\rm GeV}^2 \le  \Delta^2 \le \Delta^2_{\max}$,
where
$\Delta^2_{\max}>0$
is the limiting value
(\ref{Dmax})
of the momentum transfer squared. The left half of the graph corresponds
to the near-backward factorization regime and right half of the graph corresponds
to the  near-forward factorization (see Fig~\ref{Fig_factorization}).
With the dashed lines we show the effect of the cutoff
$\Delta^2= -1$  GeV$^2$
for the values of the CMS scattering angle.

Since these rates are certainly within the experimental reach of the
\={P}ANDA experiment, the study of reaction
(\ref{reac})
will provide a valuable universality test for the TDA approach since the same
TDAs also arise in the description of
$N \bar{N} \to \gamma^* \pi$
\cite{Lansberg:2012ha}
and backward pion electroproduction off nucleon
$\gamma^* N \to \pi N$
\cite{Lansberg:2011aa}.

It is worth mentioning that it might be advantageous also to study the process
\bea
\bar p (p_{\bar{N}})  \;+ \; n (p_N) \to  J/\psi(p_{\psi})\;+\; \pi^-(p_{\pi}).
\eea
 In our simple nucleon pole model for
 $\pi N$
 TDAs the corresponding cross sections are enhanced by the factor $2$
due to the isotopic factor
$\sqrt{2}$
in
(\ref{Nucleon_exchange_contr_VAT_pi_minus}).

\section{Conclusions}

In this Letter we address the reaction
$\bar{p} +\, N \to J/\psi \,+ \pi$
which will be studied in the  \={P}ANDA experiment at GSI-FAIR to look for exotic charmonium states production
\cite{Zcresults}.
We argue that outside the region specific for the resonance production, this reaction may be analyzed within
the pQCD framework. It will not only help to quantitatively disentangle resonance production from the universal
hadronic background but also will   provide valuable information on hadronic structure encoded in nucleon-to-pion
TDAs.

Nucleon-to-pion TDAs are essentially non-diagonal matrix elements of QCD light-cone operators which
probe the non-minimal Fock state contents of hadrons. Therefore TDAs supply complementary information
with respect to diagonal partonic  distributions (PDFs, GPDs). On the other hand, the suggested possibility
of description of the process in terms of fundamental degrees of freedom of QCD largely increases its theoretical
importance. It is also worth mentioning the possible generalization of our approach both to the case of other heavy
quarkonium  states as well as to various accompanying light meson species
($\eta$, $\rho$, $\omega$, {\it etc.}).
The TDA framework has also been recently
\cite{Goritschnig:2012vs}
used in a double handbag description of proton-antiproton annihilation into a heavy meson pair.

Within the kinematical range accessible at  \={P}ANDA we provide the predictions   using a simple nucleon
pole model for
$\pi N$
TDAs. The obtained values of cross sections give hope of experimental accessibility of the reaction. Our predictions
are consistent with the recent estimates of
\cite{Lin:2012ru}
obtained within a fully non-perturbative effective hadronic theory. However, the latter approach lacks the direct
relation to the dynamics of fundamental degrees of freedom of QCD. Precise experimental data, and the study of polarization
observables not discussed here, will allow  to discriminate between the QCD and hadronic approaches.

It is also worth mentioning that the mass of the charm quark may not be large enough
for our leading order (in
$\alpha_s$)
and leading twist analysis to be sufficient to
describe the data. More work is certainly needed to go beyond the Born approximation
for the hard amplitude, in particular because the timelike nature of the hard probe
is often accompanied by large
$O(\alpha_s)$
corrections
\cite{Muller:2012yq}.

\vskip.1in
We acknowledge useful discussions with J.P. Lansberg and are grateful to V.~L.~Chernyak for correspondence.
This work is supported in part by the Polish Grant NCN
No DEC-2011/01/D/ST2/03915,  the Joint Research Activity "Study of Strongly
Interacting Matter" (acronym HadronPhysics3, Grant 283286) under the Seventh
Framework Programme of the European Community and by the COPIN-IN2P3 Agreement.

\end{document}